# Effects of eccentricity on climates and habitability of terrestrial exoplanets around M dwarfs

Yuwei Wang[1,2], Yonggang Liu[1], Feng Tian[3,4], Yongyun Hu[1], and Yi Huang[2]

1. Laboratory for Climate and Ocean-Atmosphere Sciences, Department of Atmospheric and Oceanic Sciences, School of Physics, Peking University, Beijing, China, 100871

2. Department of Atmospheric and Oceanic Sciences, McGill University, Montreal, Quebec, Canada, H3A0B9

3. Ministry of Education Key Laboratory for Earth System Modeling, Center for Earth System, Science, Tsinghua University, Beijing, China 100084

4. Joint Center for Global Change Studies (JCGCS), Beijing, China 100875

Corresponding author: Yongyun Hu (yyhu@pku.edu.cn)

**Abstract**

Eccentricity is an important orbital parameter, which can substantially modulate the stellar insolations received by the planets. Understanding its effect on planetary climate and habitability is critical for us to search for a habitable world beyond our solar system. Planets around M dwarfs are the most promising targets due to their large population and our present-day observing technology deficiencies. The orbital configurations of M-dwarf planets are always tidally-locked at resonance states, which are quite different from those around Sun-like stars. Thus, the conclusions drawn from previous studies focusing on Sun-like-star planets may not be applicable to potentially habitable exoplanets around M dwarfs. M-dwarf planets need to be investigated separately. Here we use a comprehensive three-dimensional atmospheric general circulation model to systematically investigate how eccentricity influences climate and habitability of M-dwarf exoplanets. The simulation results show that (1) the seasonal climatic cycles of such planets are very weak even for $e = 0.4$. The global mean surface temperature varies within 5 K during an orbital cycle on an aqua planet with an Earth-like ocean (50 m oceanic mixed layer), despite a dramatic change of global mean insolation from as low as 160 W m$^{-2}$ up to as high as 800 W m$^{-2}$. Therefore, it is unlikely that an aqua planet falls out of a habitable zone during its orbit. (2) The annual global mean surface temperature significantly increases with increased eccentricity, due to the decrease of the cloud albedo. The runaway greenhouse inner edge of the habitable zone shifts outward from 2500 W m$^{-2}$ to 1700 W m$^{-2}$, and the moist greenhouse inner edge shifts from 2200 W m$^{-2}$ to 1700 W m$^{-2}$ as eccentricity increases from 0.0 to 0.4. (3) Planets in an eccentric orbit can be captured in other spin-orbit resonance states ($p$, ratio of orbital period to spin period) other than $p = 1.0$. Different spin-orbit resonance states lead to different climate patterns. Given $e = 0.4$, the climate pattern for $p = 1.0$ and 2.0 resonance states is an 'eyeball' pattern, but for half integer $p$ values such as 1.5 and 2.5, the climate pattern is a 'striped-ball' pattern, which has a belt structure with open water at low and middle latitudes and ice over both polar regions. The 'striped-ball' pattern has evidently higher surface temperatures due to the reduced planetary albedo. Correspondingly, for $e = 0.4$, the runaway greenhouse inner edge shifts outward from 1700 W m$^{-2}$ ($p = 1.0$) to 1500 W m$^{-2}$ ($p = 2.5$), and the moist greenhouse inner edge shifts outward from 1700 W m$^{-2}$ ($p = 1.0$) to 1480 W m$^{-2}$ ($p = 2.5$). Near the outer edge, planets with $p = 1.0$ and 2.0 are more resistant to the snowball state due to more locally-concentrated stellar fluxes. Thus, planets with integer spin-orbit resonance numbers have wider habitable zones than those with half-integer spin-orbit resonance states. Planets with $p = 1.0$ has the most stable climate and the widest habitable zone. As a comparison to circular orbit, eccentricity shrinks the width of the habitable zone.

# 1 Introduction

Since the first exoplanet discovered in the 1990s, more than 3600 exoplanets have been confirmed to date. One of the most intriguing goals in searching exoplanets is to find other life or even other civilizations beyond our solar system. Recent works suggested that habitable planets may be common around M dwarfs (Dressing & Charbonneau 2013; Gaidos 2013; Kopparapu 2013; Tuomi et al. 2014). Among main sequence stars, M dwarfs have the largest population. Due to limitations of present-day observational techniques, detecting exoplanets around these relatively dim stars is much easier than detecting those around other types. Therefore, exoplanets around M dwarfs are the most promising targets in searching for alien life.

Unlike planets in our solar system that all have very low orbital eccentricities and near-circular orbits, many of the discovered exoplanets have large orbital eccentricities (Wright et al. 2011; Kane et al. 2012). To our knowledge, the largest orbital eccentricity is $e$ ~0.97, which belongs to HD 20782b, a gaseous giant exoplanet (O'Toole et al. 2009). For super-Earth exoplanets, Kepler 68c has the largest eccentricity of $e = 0.42$ (Gilliland et al. 2013). It was reported that GJ 667 Cc, which is highly considered a potentially habitable exoplanet, could have an eccentricity of 0.32 (Delfosse et al. 2013). The newly discovered Proxima Centauri b also has an eccentricity (Anglada-Escude et al. 2016).

Eccentricity has important effects on exoplanets' climates and habitability (Williams & Pollard 2002; Dressing et al. 2010; Linsenmeier et al. 2015). One is that large orbital eccentricities lead to dramatic variations of stellar insolation between the periastron and apoastron. As pointed out by Dressing et al. (2010), a planet at periastron receives about twice the amount of energy at apoastron for $e = 0.2$, and this factor increases to about 9 for $e = 0.5$. In addition, the duration of the winter season around apoastron can be substantially longer than the summer season around periastron for large orbital eccentricities because a planet moves slower near apoastron than near periastron. Thus, a natural question is whether the dramatic changes in insolation would cause extreme seasonal variations of climate between apoastron and periastron. For example, if the periastron of a planet with large eccentricity is located in the habitable zone (HZ), its apoastron should be far beyond the HZ and may transit to runaway freezing, i.e., the snowball state. As the planet moves back to its periastron, it may not be able to recover. This problem has been previously studied with both energy balance models (EBM) (Dressing et al. 2010) and three-dimensional (3D) general circulation models (GCMs) (Williams & Pollard 2002, Williams & Pollard 2003, Linsenmeier et al. 2015). These works indicated that (1) extreme seasonal climate variations do not happen because the large thermal inertia of oceans acts as a buffer against dramatic insolation variations. Thus, there is no risk that large orbital eccentricities would lead to a transition to the snowball state. (2) The seasonal variations, due to eccentric orbits, can extend the outer limit of HZ to a much further distance for Earth-like planets orbiting around Sun-like stars.

While these results are important for understanding the influences of eccentricity on climate and habitability, they all focused on Earth-like planets around Sun-like stars. For potentially habitable exoplanets orbiting around M dwarfs, the situations are different. Habitable M-dwarf exoplanets are much closer to their parent stars because M dwarfs are much colder than Sun-like stars (G-type stars), so they are most likely captured in spin-orbit resonance states due to strong tidal forces (Dobrovolskis 2007). Such tidal-locking states result in distinct insolation patterns and climates. For example, if both the eccentricity and obliquity are zero, the resonance state is synchronous, and the substellar point is fixed on the equator, i.e., one side of the planets always faces their parent stars while the other side remains dark permanently. Previous works demonstrated that such insolation pattern leads to "eyeball" or "lobster" climate patterns (Pierrehumbert 2010; Yang et al. 2013; Hu & Yang 2014, Kopporapu et al. 2016), depending on whether ocean heat transports are considered. For non-zero eccentric orbits, a planet can be tidally locked to different spin-orbit resonance states other than the synchronous state (Dobrovolskis 2007, 2015). Wang et al. (2014) showed that the climate pattern is the same as that of the synchronous case, i.e., eyeball or lobster, if the resonance number is an integer, and that the climate pattern is a stripped-ball with a tropical belt of the open ocean for half-integer resonance numbers. These spatial patterns of climate are very different from that of Earth-like planets around Sun-like stars. Third, clouds have negative feedbacks in stabilizing climates of tidal-locking exoplanets around M dwarfs and can dramatically extend the inner edge of the HZ in circular orbits (Yang et al. 2013). What roles clouds play for planets in eccentric orbits have not been studied. The aim of the present paper is to systematically study the above problems for potentially habitable non-eccentric exoplanets around M dwarfs. The paper is constructed as follows. In section 2, we introduce the extended habitable zone concept. In section 3, we describe the model and experiments. In section 4, we show simulation results. The conclusion is summarized in section 5.

**2 Extended habitable zone concept**

The habitable zone is traditionally defined as a circular belt surrounding a star within which a planet can maintain liquid water on its surface. This is convenient when the orbit of a planet is circular, as it does not cross the edges of the habitable zone at any time of its orbit. A planet in eccentric orbit, however, may spend some period of time inward of the inner edge near the periastron and some period of time outside of the outer edge near the apoastron. Whether such planets are habitable needs to be clarified and the concept of the habitable zone needs to be extended to include the effect of eccentricity. The edges of traditional habitable zone are normally marked by certain stellar fluxes (Kasting et al. 1993; Yang et al. 2014; Kopporapu et al. 2016), but for a planet in eccentric orbit, the incident stellar flux varies substantially during the orbital cycle, so a representative flux needs to be chosen to give the boundaries of a habitable zone. Two reasonable choices are: (1) the mean flux over an orbital cycle or (2) the flux received when the planet is semi-major axis away from its parent star. A recent study (Bolmont et al. 2016) has tested whether we can use mean flux to indicate habitable zone. Their results showed it works well for most of the planets except for those with high eccentricity or whose parent star has high luminosity. Because the eccentricity of terrestrial planets tends to be small and M dwarfs are low-luminosity stars, the mean flux method seems to be an appropriate choice. However, the mean flux is not a quantity directly detectable by telescopes. For such reason, we prefer to use the flux received when the distance between the planet and its parent star is one semi-major axis. The advantage is that, when a planet is discovered and its semi-major axis is determined, our results are able to tell directly whether the planet is in the habitable zone for a series of possible eccentricities. Note that the two choices are equivalent: either of the fluxes and the orbital eccentricity are given, the other flux can be derived.

There are two widely used theoretical inner edges: runaway greenhouse inner edge and moist greenhouse inner edge. The former refers to where positive feedback due to water vapor greenhouse effect loses control and all surface water is evaporated, while the latter refers to where the efficient photolysis of water vapor and escape of large amounts of hydrogen into space start to occur (Kasting et al. 1993). Unfortunately, both of the inner edges are difficult to be accurately estimated in state-of-the-art 3D GCM at this stage. As for the runaway greenhouse inner edge, we need to find the point beyond which the climate system becomes physically unstable. In practice, we use the last converged model solution as a proxy for this point, following the previous studies (Yang et al. 2013, 2014; Kopporapu et al. 2016; Wang et al. 2016). However, it is hard to distinguish whether the model blows up due to physical instability or numerical instability. Wolf & Toon (2015) showed that the unconvergence of the model solution could be delayed when the deep convection component is improved. This method thus most likely underestimates the runaway greenhouse inner edge. The moist greenhouse inner edge also has large uncertainties. Accurate estimation requires the coupling of a photochemical model and a GCM model to calculate the profiles of all hydrogen bearing species, which is beyond the capability of current climate model. Some previous estimation used the water vapor content at the top of the climate model, i.e., near 3 hPa (e.g. Kopparapu et al. 2016). But certain amount of water could have already been dissociated at altitudes with greater pressure (Hu et al. 2012), thus the escape rate of water could have been underestimated. Here we simply assume that water dissociation is efficient at 100 hPa pressure level based on the previous photochemical results (Hu et al. 2012) and water starts to be lost at a significant rate when the volume mixing ratio of water vapor at this level is higher than the critical value $3 \times 10^{-3}$ (Kasting et al. 1993). The dissociation of water vapor can decrease the water vapor pressure, which facilitates the conversion of water in either liquid or solid form to water vapor. So both the liquid droplets and ice crystals in clouds are included in the estimation. To make the conclusion robust, we also push the critical level to 50 hPa as a sensitivity test.

The outer edge of the habitable zone is defined as the farthest distance at which liquid water on planetary surface is not completely frozen. Accurate estimation of the outer edge needs to deal with dense $CO_2$ atmosphere and a full description of carbon cycle in the model, which challenge the capability of today's 3D GCM. So far, all the estimations of the outer edge are estimated by one dimensional model (Kasting et al. 1993; Kopporapu et al. 2013; Wordsworth and Pierrehumbert 2013). Here we just provide some general discussions on how eccentricity may influence the outer edge for a fixed concentration of atmospheric $CO_2$, rather than accurately estimating the outer edge.

**3 Model and Experiments**

The model used in this study is the community atmosphere model version 3 (CAM3) developed in the National Center for Atmosphere Research (NCAR) (Collins et al. 2004). We used the Eulerian dynamical core at T42 spectral truncation, which is approximately $2.8°\times2.8°$ on a Gaussian grid. Hybrid $\eta$-coordinate is adopted in this version with 26 levels in the vertical. The formulation of shortwave radiation follows $\delta$-Eddington approximation of Joseph et al. (1976) and Coakley et al. (1983) and is described in detail in Briegleb (1992). The stellar spectrum is divided into 19 discrete spectral and pseudo-spectral intervals (7 intervals for $O_3$, 1 interval for the visible, 7 intervals for $H_2O$, 3 intervals for $CO_2$ and 1 interval for the near-infrared following)

(Collins, 1998). The formulation of longwave radiation is based on the broadband model approach described by Kiehl and Briegleb (1991) and Kiehl and Ramanthan (1983). Deep convection is treated with a parameterization scheme developed by Zhang and McFarlane (1995). Shallow convective overturning is treated by the parameterization of Hack (1994). Cloud fraction and the associated optical properties are evaluated via a diagnostic method in CAM3. Cloud fraction depends on relative humidity, atmospheric stability and convective mass fluxes. Three types of clouds are diagnosed by the scheme: low-level marine stratus, convective clouds, and layered clouds.

CAM3 has been extensively used in studying planetary climate (Hu & Yang 2014; Wang et al. 2014; Wang et al. 2016). In this study, CAM3 is coupled with a slab ocean, assuming the planet is an aqua planet. The depth of the slab ocean is set to be 50 m, which is the average depth of the oceanic mixed layer on Earth. The albedo of the ocean is dependent on the stellar zenith angle for direct stellar radiation (ranging from about 0.025 to 0.39), but fixed to a constant value of 0.06 for diffuse radiation. Sea ice forms thermodynamically whenever the sea surface temperature is lower than 271.35 K.

Planetary parameters are the same as those of GJ 667Cc (Anglada-Escudé et al. 2012; Delfosse et al. 2013). Planetary mass, surface gravity, and orbital period are set to be 4.27 times of the Earth's, 16.2 m s$^{-2}$, and 28 Earth days, respectively. The obliquity is assumed to be zero. The planetary atmosphere is assumed to contain 1 bar of background gas (e.g., $N_2$). $CO_2$ concentration is fixed at 355 ppmv in all simulations. The stellar spectrum at the top of the atmosphere (TOA) is generated from a 3700 K Planck function, which peaks at 0.78 μm.

GJ 667Cc is a highly potentially habitable terrestrial exoplanet around M dwarf, receiving about 1237 W m$^{-2}$ stellar flux. Its parent star GJ 667C is a member of the Gliese 667 trinary star system, with GJ 667 A and B both being more massive than GJ 667C (Anglada-Escudè et al. 2012, Delfosse et al. 2013).This planet is our neighbor, located only 23.6 light years (7.2 pc) away, making it one of the most popular candidates for searching for alien life. However, the eccentricity of GJ 667Cc still has some uncertainties. It was initially reported to have an eccentricity lower than 0.27 (Anglada-Escudè et al. 2012). Another group later announced that its eccentricity is about 0.32 (Delfosse et al. 2013). Then, the first group corrected the eccentricity to be near-zero (Anglada-Escudè et al. 2013). Theoretical numerical simulations indicated that the orbital eccentricity of GJ 667Cc changes cyclically in the range 0.05 - 0.25 with a period of approximately 0.46 Earth year (Makarov et al. 2013) and further pointed out that GJ 667Cc is likely (probability = 0.51) to be entrapped in $p = 1.5$ (refer section 4.1 for definition of $p$) or even higher spin-orbit resonance state. Thus, it is particularly interesting to have GJ 667Cc as an example to study how eccentricity affects the climate and habitability of potentially habitable exoplanets around M dwarfs.

## 4 Results

### 4.1 Insolations received by M–dwarf planets in eccentric orbit

Compared to planets in circular orbit, planets in eccentric orbit receives more flux by a factor of $(1-e^2)^{-\frac{1}{2}}$, given the same semi-major axis. Figure 1 schematically shows the comparison between the small- and the large-eccentricity orbit. In circular orbit, the annual mean stellar flux received by GJ 667Cc at the substellar point is approximately 1237 W m$^{-2}$. If the eccentricity increases to 0.32, the annual mean stellar flux rises to $1237 \times (1-0.32^2)^{-\frac{1}{2}} = 1305$ W m$^{-2}$, an increase of 68 W m$^{-2}$ or 5.5 percent.

Another difference from circular orbit is the movement of the substellar point. Rather than fixed on the planet, the substellar point swings zonally back and forth along the equator. In a circular orbit, the orbital angular velocity is constant and equals to the spin angular velocity. Thus, the substellar point is relatively stationary to the planet. In an eccentric orbit, according to Kepler's second law, the orbital angular velocity is larger near the periastron and smaller near the apoastron. The speed and direction of the movement are determined by the difference between orbital angular velocity and spin angular velocity.

When the eccentricity is small, the orbital period is equal to the spin period. As eccentricity increases, the planets are more likely captured into higher spin-orbit resonance states. A spin-orbit resonance number is commonly defined as:

$$p = \frac{Average\ spin\ angular\ velocity}{Average\ orbital\ angular\ velocity} = \frac{Rotation\ period}{Spin\ period}.$$

*p* equals to 1.0 when the rotation period is the same as the spin period. Mercury spins three times when it completes two rotations around the Sun, so its resonance number *p* is 1.5. For close-in planets with zero eccentricity, the probability of planets being captured in *p* = 1.0 state is 100%. This is why we always assume that the rotation period equals to the spin period for M-dwarf planets in circular orbits. The probability of being captured in certain spin-orbit resonance number is also sensitive to planets' initial spin velocities. According to constant Q-tides model prediction, for rapidly spin planets with *e* = 0.4, the most probable spin-orbit resonance numbers are *p* = 2.0 and 2.5, with the probability to be approximately 33% for both of them. If the initial spin is slow, the most probable spin-orbit resonance number is *p* = 1.5, with a probability of ~55% (Dobrovolskis 2007). Different spin-orbit resonance numbers, of course, lead to different insolations. In our previous paper (Wang et al. 2014) discussing different climate patterns associated with different resonance numbers, we inappropriately round the spin period (days) to an integer, which makes our results uncorrect. (Dobrovolskis 2015). We correct this mistake in section 4.4.1 in this paper.

**4.2 Seasonal cycle**

For Earth, the seasonal cycle is caused by the obliquity because of its near-zero eccentricity. Obliquity changes the spatial distribution of stellar insolation without changing the global mean, while eccentricity changes both spatial distribution and global mean stellar radiation. In the present paper, because the obliquity is assumed zero, the seasonal cycle is solely due to the eccentricity.

In an eccentric orbit, a planet receives the highest stellar radiation at the periastron and the least stellar radiation at the apoastron (Figure 1). In this section, we assume the stellar flux received by the planet at the semi-major axis is 1237 W m$^{-2}$, and the orbital period is equal to the rotation period, i.e., $p = 1.0$. Figure 2a shows the annual cycles of the global mean insolations for three different eccentricities. For $e = 0.0$, the insolation is a constant 309.25 W m$^{-2}$. For $e = 0.2$, the planet receives ~480 W m$^{-2}$ at periastron at day 0 and day 27, and ~215 W m$^{-2}$ at the apoastron at day 13 and day 14. When eccentricity increases to 0.4, the planet receives more than 800 W m$^{-2}$ at the periastron and less than 160 W m$^{-2}$ at the apoastron. As a comparison, the annual mean insolation of the Venus, the Earth, and the Mars are approximately 660, 341, and 148W m$^{-2}$, respectively. Because the insolations of large-eccentricity planets have dramatic swings over an orbital cycle, one may expect those surface temperatures of these planets would also have large seasonal cycles. However, our simulation results show that the seasonal cycles of surface climate are very weak. Even for eccentricity $e = 0.4$ (Figure 2b), the seasonal variation of global mean surface temperature is less than 5 K. This result is consistent with a previous study on an Earth-like planet (Williams & Pollard 2002). They showed that the seasonal cycle for an Earth-like planet is less than 15 K at $e = 0.4$. The larger amplitude (15 K) is a result of the Earth's longer orbital period (365 days).

The weak seasonal cycles of surface temperatures are owing to the large thermal inertia of the ocean mixed layer. The mixed layers of Earth's oceans, which are on top of the oceans, can be as thin as a few meters or as deep as 2000 m. It is about 50 m on global average. Due to active turbulent mixing, the properties of ocean water are nearly vertically uniform within the mixed layer. Changes in ocean surface temperatures are closely related to that of the whole mixed layer. A small change (increase or decrease) in ocean surface temperature requires a large amount of change in energy (absorbed or released). Figure 3a shows the seasonal variations of surface temperatures averaged over the permanent ocean area (not including the region which is fully or partially covered by sea ice) from GCM simulations for $e = 0.4$. For 50 m Earth-like mixed layer, the seasonal variation of ocean surface temperature is merely 1 K. As a sensitivity test, we decrease mixed layer depth to extremely low 10 m, the seasonal variation is still limited to less than 3.6 K.

To further explain the mechanisms on how ocean mixed layer stabilizes ocean surface temperature, we use an idealized mixed-layer model (Pierrehumbert 2010, Chapter 7.4 ). The change of ocean surface temperatures can be expressed as:

$$T'(t) = \frac{1}{\rho c_p H} \int_0^t (1 - \alpha) S'(t') dt'. \quad (1)$$

Here, $T'(t)$ and $S'(t)$ are the changes of ocean surface temperatures and stellar insolations relative to the annual mean values, respectively. $\rho$ is the density of the ocean water, and $H$ is the depth of the ocean mixed layer. Thus, $\rho H$ represents the mass over a unit area. $C_p$ is the heat capacity. The absorption and scattering of stellar fluxes by the atmosphere and clouds as well as surface albedo are all included in $\alpha$. Based on the values diagnosed from GCM simulations, $\alpha$ is set to 0.7. $S'(t)$ is also diagnosed from GCM simulations (Figure 3b). It is found that $S'(t)$ changes dramatically, ranging from ~-340 W m$^{-2}$ to ~1220 W m$^{-2}$. When $S'(t)$ is applied to the idealized model (equ. (1)), results of both seasonal amplitude and phase of ocean surface temperature are consistent with those in GCM (Figure 3c). The seasonal variations of surface temperature are 0.65 K and 3.4 K for 50 m and 10 m ocean mixed-layers respectively. The hottest day is about 4 days after the planet passes the periastron, and the coldest day is about 10 days behind the apoastron (Figure 3a & 3c), when $S'(t)$ becomes zero (Figure 3b). The asymmetry of lag time is due to the faster orbital angular velocity near periastron and slower angular velocity near apoastron.

The sensitivity of temperature seasonal variations to orbital periods is also tested using GCM (Figure 3d). In these tests, the mixed-layer depth is fixed at 50 m. It is found that the amplitude of the seasonal variations increases with the orbital period. The amplitude is ~0.6 K, ~1.5 K, and ~3.4 K for orbital period of 20, 40, and 80 days. Based on the diagnosed $S'(t)$ (Figure 3e), the idealized mixed-layer model predicts seasonal variations of 0.47, 0.96, and 1.93 K for the three orbital periods. The differences between the idealized model and GCM are mainly due to different planetary albedos, which are all fixed at $\alpha = 0.7$ in the idealized model. The above results indicate that the seasonal temperature variations of M-dwarf habitable planets are weak, as long as there are large open oceans on the planets. The large thermal initial of ocean is able to damp out dramatic insolation changes caused by eccentricity and stabilize planetary climates. Although the above results are only for the $p = 1.0$ resonance state, they can also be applied to other spin-orbit resonance states.

**4.3 Climate and habitability at $p = 1.0$ spin-orbit resonance state**

It has been recently shown that eccentricities of terrestrial planets tend to be small (Van Eylen & Albrecht 2015). Combined with the relationship between eccentricity and spin-orbit resonance numbers as described above, we can infer that a large part of terrestrial planets around M dwarfs should be captured in small spin-orbit resonance states, especially $p = 1.0$ state. Therefore, we pay more attention to the climate and habitability of planets in $p = 1.0$ state. As the largest eccentricity of terrestrial planet that has been observed so far is 0.42, we limit the eccentricity in this study to 0.4.

**4.3.1 Climate at 1237 W m$^{-2}$**

As stated before, 1237 W m$^{-2}$ is the insolation received by GJ 667Cc at its semi-major axis. Figure 4 shows the annual mean surface temperature distributions from $e = 0.0$ to $e = 0.4$. The black curves are the isotherms of the freezing point (271.35 K), i.e., the sea ice boundaries. The black arrows show the ranges of substellar points' migration. The substellar point is fixed at 180°E (black dot) at $e = 0$. As the eccentricity increases, the surface temperature increases almost everywhere and the open ocean extends more towards the nightside. The global average surface temperature increases by 21 K, from 236 K at $e = 0.0$ to 257 K at $e = 0.4$ (Figure 5a), and the open ocean expands from 30% of the planet surface at $e = 0.0$ to 45% at $e = 0.4$ (Figure 5b).

Two types of changes are involved that may contribute to the warming: (1) the average distance between the planet and the star decreases (Figure 1), therefore the annual mean stellar flux received by the planet increases which directly causes the warming. (2) The shape of the orbital changes and the associated insolation pattern changes may trigger some internal adjustments in the climate system that cause the warming. As will be proved later with numerical simulations, these two types of changes are almost orthogonal. In order to distinguish and quantify the influences of these two types of changes, we design two additional sets of experiments and refer to the original experiments as the 'Control' experiments. In the first set of experiments (referred to as the 'Flux Change' experiments), the orbit is fixed to be circular but the stellar flux at TOA is varied. Recall that the annual mean insolation increases with $e$ by a factor of $(1-e^2)^{-\frac{1}{2}}$. The insolations are set to 1237 W m$^{-2}$ ($e = 0.0$), 1243 W m$^{-2}$ ($e = 0.1$), 1263 W m$^{-2}$ ($e = 0.2$), 1298 W m$^{-2}$ ($e = 0.3$) and 1350 W m$^{-2}$ ($e = 0.4$), respectively. In the second set of experiments (referred to as the 'Pattern Change' experiments), the orbital shape is changed based on the eccentricity but the annual mean insolation is fixed to be 1237 W m$^{-2}$.

The results are shown in Figure 5. Their sum curves (black dotted) resemble control experiment curves (red), which proves our orthogonal assumption. Surface temperature increases by 21 K as eccentricity increases from $e = 0$ to $e = 0.4$, approximately 15 K due to internal changes of climate system induced by the change of orbital shape, and only 5 K due to the changes of the total insolation (Figure 5a). Similar to the surface temperature, the internal changes contribute 0.11 and insolation changes contribute 0.02 to the total ice fraction change of 0.15 in control experiments. The small discrepancy in sea-ice fraction between the sum of the two factors acting individually and sea-ice fraction obtained in control experiment where two factors act simultaneously is due to the nonlinearity of the climate system: the two factors reinforce each other slightly when acting together. Overall, the result demonstrates that the change of the planet's orbital shape contributes much more than the change of mean distance (i.e. mean stellar flux) to the climate change when eccentricity changes.

In 'Pattern change' experiments, since the concentration of greenhouse gas is fixed, the warming of the climate is caused by the reduction of the planetary albedo. The planetary, surface and cloud albedos are calculated and shown in Figure 6a. The planetary albedo decreases from 0.47 at *e* = 0.0 monotonically to 0.4 at *e* = 0.4. The decrease of planetary albedo is related to the migration of the substellar point, which is the major consequence when the shape of the planetary orbit becomes none-circular. The substellar point oscillates around the center (180°E) of the dayside along the equator (Figure 4 and Figure 7a). The larger the eccentricity is, the farther the substellar point oscillates away from the center of the dayside.

Planetary albedo is mainly determined by cloud properties and surface characteristics for Earth-like atmosphere. Rayleigh scattering by atmospheric molecules is much weaker when the surface atmospheric pressure is fixed to be 1 bar (Wolf & Toon 2015). Figure 6a shows that the decrease of the planetary albedo here is fully due to the decrease of cloud albedo. Further inspection shows that the cloud albedo is lowest around day 4 and day 24 and highest around day 14 and day 0 or day 27 for *e* = 0.4 (Figure 7b). Day 4 and day 24 are the times when the substellar points are at the westernmost and easternmost positions (Figure 7a). The surface temperatures at these two locations are much lower than that in the central region (Figure 4). Recall that we have proved the weak seasonal cycle, so such temperature gradient always persists even when the star directly shines on the westernmost or easternmost positions. Figure 7c shows the surface temperature under the substellar point (averaged over a 60° circular disk around the substellar point). The temperatures at day 4 and day 24 are only about 286 K and 284 K, respectively, while it can reach as high as 292 K at day 14 and day 0 or 27 for *e* = 0.4. The difference is more evident as eccentricity increases since the substellar point can move farther from the central region. Note that this feature is not dependent on the size of the circular disk chosen.

The surface temperature influences the cloud albedo through convection. Because of the slow rotation rate, the free atmosphere is in weak temperature gradient regime (Showman et al. 2013), which means that the temperature is nearly uniform in the free atmosphere. So the vertical temperature gradient, which is important to atmospheric stability, is much smaller over regions where the surface temperature is low than that over regions where the surface temperature is high. The convective available potential energy (CAPE) is smaller and convective inhibition (CIN) is larger over low surface temperature regions than those over central region. Besides, because of the surface temperature gradient, the horizontal winds near the surface converge in the central region in climatology. It provides strong positive forcing to the convection. The larger the surface temperature gradient is, the stronger the convergence becomes. Vertical velocity at 500 hPa under substellar point (averaged within a 60° disk around the substellar point) (Figure 7d) supports this argument. The upward motion, or convection, is weakest around day 6 and day 23, and strongest around day 14 and day 28. Again, the size of the disk does not affect the characteristics of the result. Due to the slow spin rate, convection-driven clouds are prevalent, while wave-driven clouds prevailing on Earth's mid-latitude regions are near absent on this kind of planet. Thus, the total cloud water averaged within 60° circular disk around the substellar point is most abundant near day 14 and day 0 or 27, and least abundant near day 4 and day 22 (Figure 7e). Therefore, the weaker convection in the colder region is responsible for the decrease of the cloud albedo.

Besides the decrease of the planetary albedo, weak convection brings in another effect. Convection is an efficient way of transporting heat from the surface to the upper atmosphere. Weaker convection in large-eccentricity simulations therefore reduces this efficiency. This effect is clearly visible in Figure 6b, which shows the atmospheric temperature profiles over the permanent ocean region for different eccentricities. Near the surface, air temperature at $e = 0.4$ is 2.5 K higher than that at $e = 0.0$, but the relation is reversed above 600 hPa, air temperature at $e = 0$ exceeds that at $e = 0.4$. In other words, weak convection leads to a larger atmospheric lapse rate, which enhances the surface warming as eccentricity increases.

### 4.3.2 Climates and habitability at other insolations

When insolation (at one semi-major axis distance) increases or decreases, the global average annual mean surface temperature increases or decreases almost linearly.

The larger the eccentricity is, the faster the surface temperatures change, suggesting larger climate sensitivity to insolation for larger eccentricities (Figure 8a). Because of the different slopes of the trends, the temperatures for different eccentricities converge as the insolation decreases. In fact, the temperatures are nearly the same for all eccentricities when insolation is below 600 W m$^{-2}$. The trend for global ice fraction is similar (Figure 8b).

Two methods of calculating the inner edges of the habitable zone have been introduced in section 2, and both of them give the result that the inner edge moves outward significantly with increased eccentricity. The runaway greenhouse inner edge is 2500 W m$^{-2}$ for $e = 0.0$, decreases to 2400 W m$^{-2}$, 2200 W m$^{-2}$, 1900 W m$^{-2}$ and 1700 W m$^{-2}$ for $e = 0.1, 0.2, 0.3$ and $0.4$, respectively (Figure 8e). The moist greenhouse inner edge moves outward slightly less (Figure 8c). The black dashed line in Figure 8c shows the critical value of volume mixing ratio of water vapor, $3 \times 10^{-3}$. It can be seen that the moist greenhouse inner edge is approximately 2200 W m$^{-2}$ for both $e = 0.0$ and $e = 0.1$, decreases to 2100 W m$^{-2}$, 1900 W m$^{-2}$ and 1700 W m$^{-2}$ for $e = 0.2, 0.3$ and $0.4$, respectively (Figure 8e). Results change only slightly if the critical escape level is assumed to be 50 hPa (Figure 8d).

### 4.4 Climates and habitability for other spin-orbit resonance states

As mentioned in section 4.1, planets can be captured in higher spin-orbit resonance states ($p > 1.0$) when eccentricity increases. The most probable resonance numbers for $e = 0.4$ are $p = 1.5, 2.0$ and $2.5$ (Dobrovolskis 2007). In this section, we focus on how these different spin-orbit resonance states affect planetary climate and habitability.

#### 4.4.1 Climates at insolation of 1237 W m$^{-2}$

Again, we start with the insolation (at the semi-major axis) of 1237 W m$^{-2}$. The eccentricity is fixed to 0.4. Figure 9 shows the annual mean insolation distributions for

$p$ = 1.0, 1.5, 2.0 and 2.5 states. There are two different kinds of insolation patterns. For integer $p$ values, e.g. 1.0 and 2.0, the insolation has one maximum center. Most stellar fluxes are concentrated in a limited area on one side of the planet, and very little fluxes are irradiated into the other side. Note the reason for such pattern appearing for $p$ = 2.0 is that the distance between planet and its parent star is always farthest when the ice side (centered at longitude 0°E) is facing the star. For half-integer resonance states $p$ = 1.5 and 2.5, there are notably two centers of insolation, one on each side of the planet. The intensities of the insolations at two centers are the same. Such insolation patterns are consistent with those in Dobrovolskis (2015).

Figure 10 shows the distributions of annual-mean surface temperatures. The black curves are again the isotherms of the freezing point (271.35 K), i.e., the sea ice boundaries. The patterns of surface temperature somehow resemble those of the insolations, but with some differences due to the atmospheric dynamics. For the cases of $p$ = 1.0 and 2.0, surface temperature shows classic eyeball pattern (Pierrehumbert 2010), a limited near-circular region of open water on one side of the planet and ice on the rest of the planet. For those of $p$ = 1.5 and 2.5, a different pattern appears in which a global belt of open water exists in the low and middle latitudes and ice over both polar regions. In Wang et al. (2014), although we failed to get the right insolation pattern, the climate pattern only has a small difference with the pattern we obtained here. The sea ice boundaries are wave lines rather than straight lines. Thus, we still call this pattern 'striped-ball' pattern following Wang et al. (2014). This pattern is also shown in the simulations of Proxima Centauri b (Boutle et al. 2017). This belt structure, different from insolation's double eyeball structure, is largely owing to the atmospheric heat transport that greatly reduces the zonal temperature gradient. The temperature gradient would otherwise be very large because of the large gradient of insolation. The largest insolations (two bright centers in Figures 9b and 9d) are higher than 700 W m$^{-2}$ while the smallest values (in the middle between the two bright centers in Figures 9b and 9d) can be less than 300 W m$^{-2}$. Assuming 0.7 planetary albedo, radiative balance equations yield a surface temperature difference as large as 55 K. By contrast, the largest temperature difference around the equator obtained by GCM simulations is only within 10 K for $p = 1.5$ and 5 K for $p = 2.5$. Further tests show that the climate always evolves into one of the two climate patterns no matter we start the model from an ice-free or full-ice initial conditions.

The striped-ball climate is much warmer than the eyeball climate. The global mean surface temperatures of the eyeball climate are 257 K and 261 K for $p = 1.0$ and $p = 2.0$, but 277 K and 281 K for $p = 1.5$ and 2.5 (Figure 11a). The global ice fraction is also higher for the eyeball climate than for the striped-ball. The eyeball climate has global ice fraction of ~56%, while the stripped-ball has ~22% (Figure 11b). The two climate patterns have ~20 K difference in the global mean temperature and ~35% difference in global ice fraction.

Again we start by looking at planetary albedos for clues why the temperatures are different between different $p$ values. Figure 11c shows that the planetary albedos for integer $p$ (eyeball pattern climate) are generally higher than those for half-integer $p$ (striped-ball pattern climate). The difference in planetary albedos is not solely due to the difference in cloud albedos as was the case described in section 4.1 (Figure 11d), but also due to the surface albedo. The cloud albedos and the mechanism identified in section 4.1 still play major roles for cases $p = 1.0$, 1.5 and 2.5, but not for the case $p = 2.0$ (third row of Figure 12).

For $p = 1.0$, 1.5 and 2.5, surface albedos are stable. The variations are around 0.1, very small fluctuations, and the surface albedo is much smaller than the cloud albedo (Figure 12). In contrast, for $p = 2.0$, surface albedo fluctuates with amplitudes comparable to that of the cloud albedo. The reason is that for $p = 2.0$, the substellar point has to move through the entire ice-covered region in every spin period around day 14 (Figure 12c and 12g). During those times, the surface albedo is high but the cloud albedo is low, and the high surface albedo compensates the low cloud albedo induced by weak convection over ice, thus the resulting planetary albedo is similar to that of the $p = 1.0$ case. Interestingly, although the substellar point goes through the entire planet, only one side of the planet exists an open ocean. Combined with insolations patterns

(Figure 9), the ice side cannot receive enough energy to melt the ice and sustain the open ocean. Readers can refer the detailed explanations in Wang et al. (2014). For $p = 1.0$ case, the substellar point moves only within a narrow range of longitude and does not really move into the ice covered region; for both $p = 1.5$ and 2.5 cases, there is no sea ice in the tropical region (Figures 10b and 10d), therefore the substellar point is never in the ice covered region although it moves through the whole equatorial circle. We have to emphasis that the surface temperature and sea ice can interact with each other. Once the surface temperature increases, the sea ice starts to melt which lowers the surface albedo, and in turn increases the surface temperature. This is the so-called 'ice-albedo' feedback. This is why we got two quite different climate patterns: one has low global mean surface temperature and large ice fraction, and the other has high global mean surface temperature and low ice fraction.

The cloud albedos still correlate well with the surface temperatures under substellar point regions for all cases. The panels on the right of Figure 12 show the average surface temperatures in $60^o$ disk around the substellar point. Note that the cloud albedo is most related to the cloud under the substellar point region. The cloud in night side no matter dense or thin has no relationship with the cloud albedo. That's why we choose a disk around the substellar point region, not the global, to calculate the average surface temperature for comparison with cloud albedo. Comparison (middle and right columns of Figure 12) clearly shows that high cloud albedo is always obtained at higher surface temperatures, which is consistent with what we have discussed in section 4.1. The cloud albedo of $p = 1.0$ is higher than that of $p = 1.5$, and the cloud albedo of $p =1 .5$ is higher than that of $p =2.5$. There are two reasons here: surface temperature gradient and spin rate. Surface temperature gradient in low $p$ values are larger, so the horizontal winds near the surface are stronger, leading to more intense convergence in the hot spot. Spin the rate can influence the distribution of cloud. Fast-spin planets always have lower cloud albedos than slow-spin planets (Kopporapu et al. 2016).

**4.4.2 Climates and habitability at other insolations**

Climates for a series of insolations mimicking the different lengths of the semi-major axes are calculated in order to determine the width of the habitable zone. Similar with what we have found in section 4.2, the surface temperatures increase almost linearly with the insolations for none-unique $p$ values too (Figure 13a). On average, $p = 2.5$ has the largest slope, indicating that its surface temperature increases fastest if the planet is moved closer to the parent star, followed by $p = 1.5$, 2.0 and 1.0.

When the last converged solution is taken as the runway greenhouse inner edge, $p = 1.0$ has the innermost one with value approximately 1700 W m$^{-2}$. The edges for $p = 1.5$ and 2.0 are close to each other, with values 1550 W m$^{-2}$ and 1600 W m$^{-2}$ respectively. The runaway greenhouse inner edge for $p = 2.5$ is the outermost, ~1500 W m$^{-2}$ (Figure 13e). Figure 13c shows the mixing ratio of water at 100 hPa. The shift of the moist greenhouse inner edge is similar to that of the runaway greenhouse inner edge. The volume water mixing ratio of $p = 2.5$ is the first to reach the criterion of $3\times10^{-3}$ (black dashed line), at insolation ~1480 W m$^{-2}$, followed by $p = 1.5$ and 2.0, both at insolation ~1550 W m$^{-2}$. The moist greenhouse inner edge for $p = 1.0$ is still the innermost, at ~1700 W m$^{-2}$. These are all summarized in Figure 13e. Results are similar if the critical escape level is assumed to be 50 hPa (Figure 13d).

As the insolation decreases to ~700 W m$^{-2}$, climates for $p = 1.5$ and 2.5 first transit into snowball state (Figure 13b). For the cases of $p = 1.0$ and 2.0, the insolation needed for snowball formation have to decrease further to ~500 W m$^{-2}$. The reason why $p = 1.0$ and 2.0 are more resistant to being globally frozen is that most of the incident fluxes are concentrated on one side of the planet (Figures 9a and 9c). The concentrated fluxes make it much easier for the planet to keep an open ocean.

These analyses reveal that the climate of planets with $p = 1.0$ is the most stable compared to planets with other spin-orbit resonance numbers. Planets with integer spin-orbit resonance numbers have wider habitable zone than planets with half-integer spin-orbit resonance numbers.

## 4.5 Dependence of climates on spin period in eccentric orbit

Recent works suggested fast spin planets have higher surface temperatures and more outward runaway greenhouse inner edges than slow spin planets (Yang et al. 2014; Kopparapu et al. 2016). In order to investigate what role the spin period plays in an eccentric orbit, we try four rotation periods, 10 days, 20 days, 40 days and 80 days, for selected spin-orbit resonance numbers and eccentricities. The selected configurations are $p = 1.0$ with $e = 0$ and $p = 1.0, 1.5, 2.0$ and $2.5$ with $e = 0.4$, therefore 20 simulations are carried out in total. The corresponding spin period with certain resonance number can be determined exactly by rotation period and $p$. The insolations at semi-major-axis distance used in all simulations are 1237 W m$^{-2}$.

Figure 14a shows the global mean surface temperatures for all configurations. For the circular orbit ($e = 0.0$, $p = 1.0$), the surface temperature decreases from 249 K to 237 K as rotation period as well as spin period increase from 10 days to 20 days (Figure 14a), consistent with previous study (Kopparapu et al. 2016). The decrease of surface temperature is due to the increase of the cloud albedo from 0.27 to 0.36 (Figure 14b). Still for $p = 1.0$, when $e = 0.4$, the surface temperature also decreases from 264 K to 256 K as rotation period increases from 10 days to 20 days (Figure 14a), and the cloud albedo increases from 0.24 to 0.3 (Figure 14b). Climate patterns of other spin-orbit resonance states, especially for $p = 1.5$ and $2.5$, may be different from that of $p = 1.0$ state (Figure 10), but they all show cooling of the surface temperature with increasing rotational periods (Figure 14a). All results can be explained by the increase of the cloud albedos (Figure 14b). Thus, the eccentricity does not change the dependence of climates on spin period, no matter what spin-orbit resonance state the planet is captured in.

## 5 Conclusions

In this study, we use a comprehensive 3D GCM to systematically investigate how eccentricity (from 0.0 to 0.4), combined with different spin-orbit resonance states and rotation periods, affects the climate and habitability of planets around M dwarfs. When varying the eccentricity, we have chosen to fix the stellar flux received by planet when the planet is semi-major axis away from its parent star. Our conclusions are as follows:

(1) The large thermal inertial of oceans can efficiently damp out the large variations of stellar insolations. The seasonal cycles of M-dwarf planets with highly eccentric orbits are very weak, as opposed to dramatically changing insolations. Taking the case of $e = 0.4$ as an example, the global mean surface temperature of an aqua planet with Earth-like ocean (50 m oceanic mixed layer) varies by less than 5 K during one orbit, although the stellar flux at the top of the atmosphere varies from 160 W m$^{-2}$ at apoastron to 800 W m$^{-2}$ at periastron. The seasonal cycle remains weak even when the oceanic mixed-layer depth is reduced to 10 m or the rotational period extended to 100 days.

(2) Based on the theoretical model predictions and observational data, most M-dwarf exoplanets are in $p = 1.0$ spin-orbit resonance state, i.e., the spin period is equal to the rotation period. For such resonance state, the global mean surface temperature is found to increase with increased eccentricities. Assuming the insolation at semi-major axis to be 1237 W m$^{-2}$, the global mean surface temperature increases by 21 K as the eccentricity increases from 0.0 to 0.4. The primary cause of this warming is the farther movement of the substellar point to colder regions, where convective strength is weaker and the cloud and planetary albedos are smaller. A secondary cause is the increased annual-mean stellar insolation, which explains a quarter of the warming. Correspondingly, the runaway greenhouse inner edge of the habitable zone shifts outward significantly from 2500 W m$^{-2}$ to 1700 W m$^{-2}$ when the eccentricity increases from 0.0 to 0.4. The moist greenhouse inner edge shifts less from 2200 W m$^{-2}$ to 1700 W m$^{-2}$ for the same change of eccentricity.

(3) Eccentricity may render the planets captured in spin-orbit resonance states other than $p = 1.0$ state. Different spin-orbit resonance states are found to have influences on climates. Given $e = 0.4$, the climate pattern for $p = 1.0$ and 2.0 resonance states is 'eyeball' pattern, but for $p = 1.5$ and 2.5 resonance states the climate pattern is 'striped-ball' pattern in which a global belt of open water exists at low and middle latitude and two ice caps cover the polar regions. This striped-ball pattern has higher surface temperatures due to the reduced planetary albedos. Correspondingly, for $e = 0.4$, the runaway greenhouse inner edge shifts outward from 1700 W m$^{-2}$ ($p = 1.0$) to 1500 W m$^{-2}$ ($p = 2.5$), and the moist greenhouse inner edge shifts outward from 1700 W m$^{-2}$ to 1480 W m$^{-2}$. Near the outer edge, planets with $p = 1.0$ and 2.0 are more resistant to snowball glaciation than those with $p = 1.5$ and 2.5. In summary, planets in $p = 1.0$ resonance state have the most stable climate and the widest habitable zone. Planets with integer spin-orbit resonance states have wider habitable zone than these with half-integer spin-orbit resonance states.

(4) Compared to the circular orbit, eccentricity shrinks the width of the habitable zone.

**Acknowledgment:** Y. Wang and Y. Hu are supported by the National Natural Science Foundation of China under grants 41375072 and 41530423. Y. Liu is supported by the Startup Fund of the Ministry of Education of China. F. Tian is supported by the National Natural Science Foundation of China (41175039) and the Startup Fund of the Ministry of Education of China. Y. Huang acknowledges the grants from the Natural Sciences and Engineering Research Council of Canada (RGPIN 418305-13) and the Fonds de recherche du Québec (2016-PR-190145).

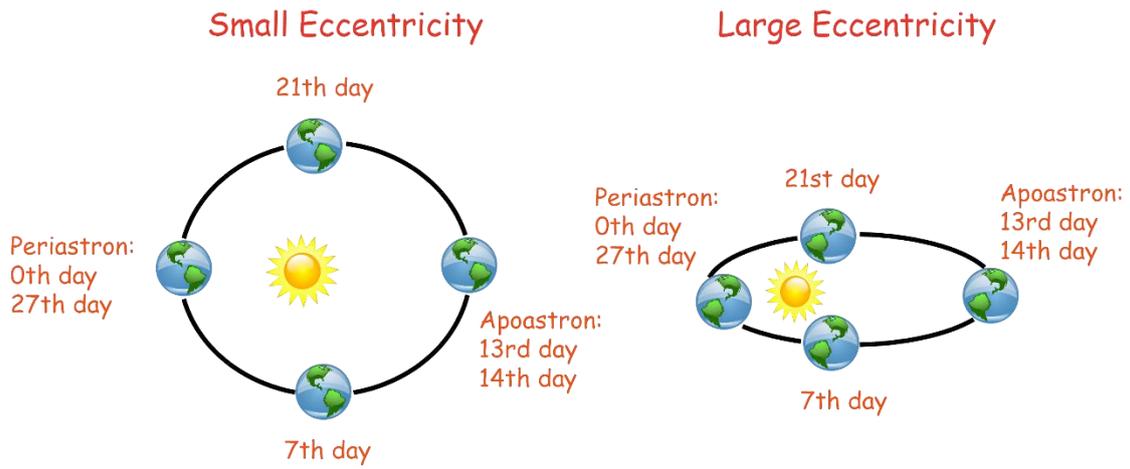

**Figure 1.** Schematic diagram of small-eccentricity and large-eccentricity orbits. The semi-major axises of these two orbits are the same, and therefore the orbital periods are the same, 28 days. Planets move past the periastron at the end of the 27$^{th}$ day or the beginning of the 0$^{th}$ day, and past the apoastron at the end of the 13$^{rd}$ day or the beginning of the 14$^{th}$ day..

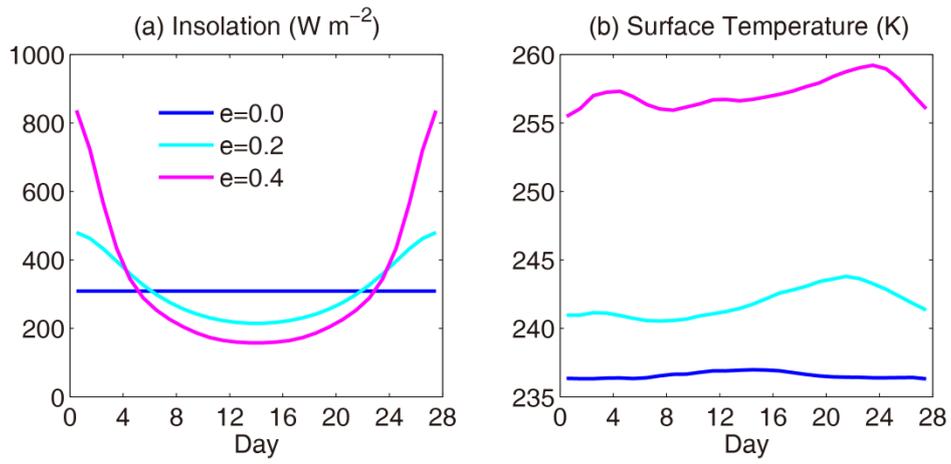

**Figure 2.** Variations of (a) global mean insolations and (b) surface temperatures in one orbital cycle for different eccentricities. The oceanic mixed layer depth is 50 m.

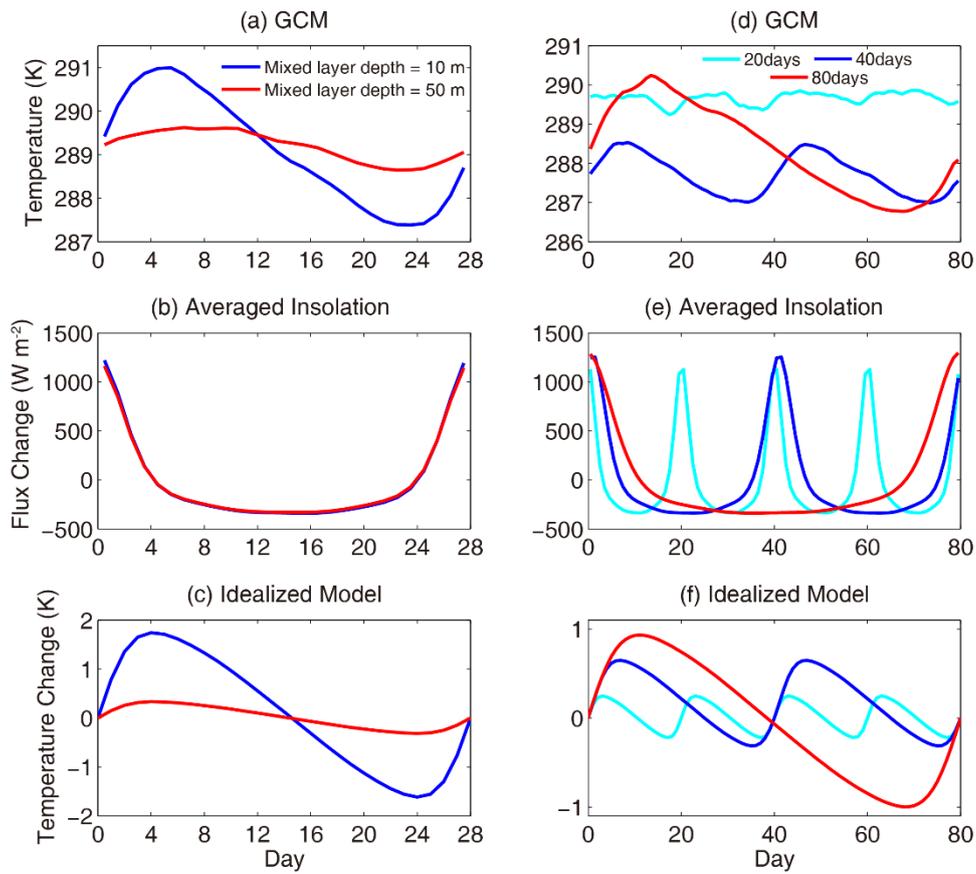

**Figure 3.** Variations of (a) surface temperatures and (b) insolations averaged over permanent open ocean regions for two different oceanic mixed layer depths in GCM. The orbital period is fixed to 28 days. (c) Oceanic temperature changes predicted by the idealized model (equ. (1)). (d) – (f) The same as (a) – (c) but for different orbital periods with fixed 50 m oceanic mixed layer.

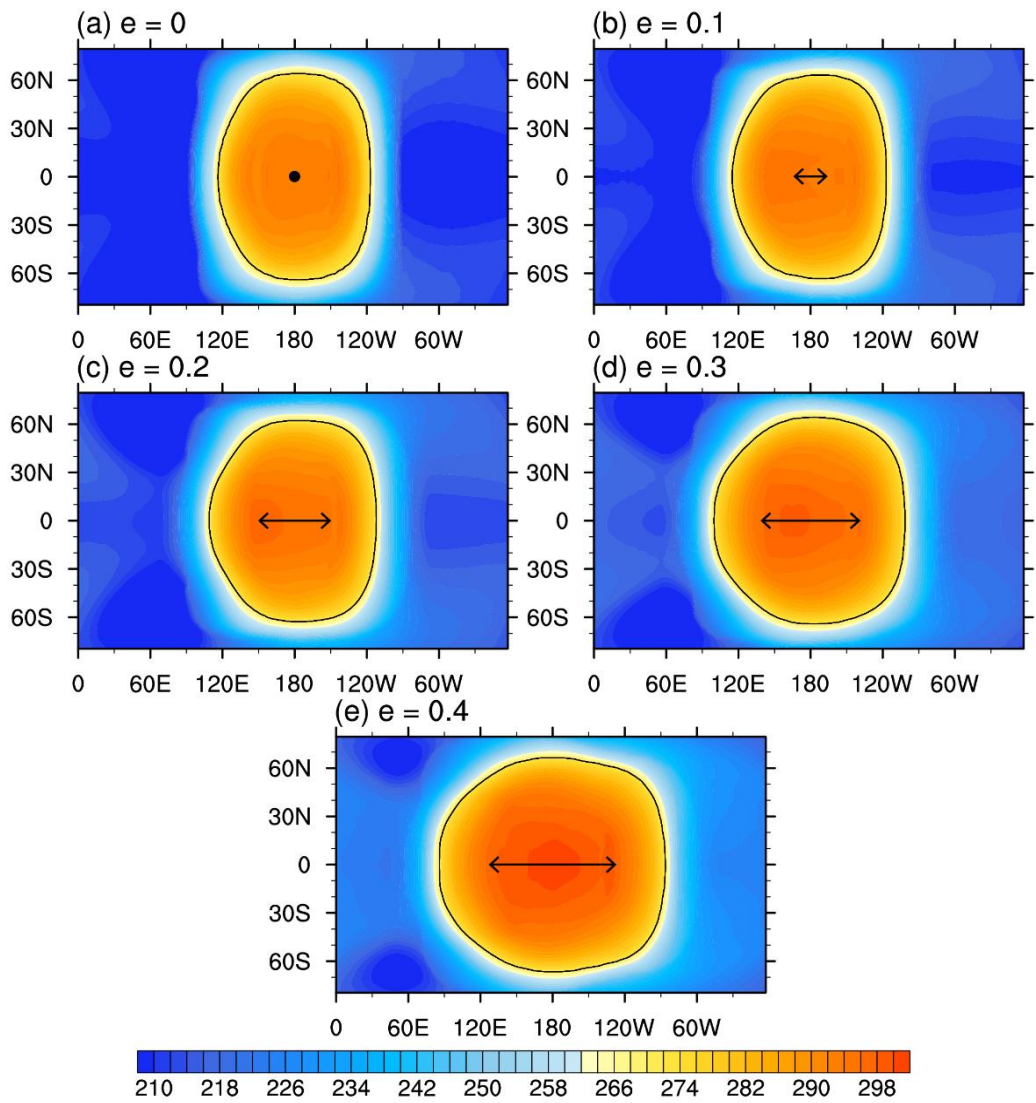

**Figure 4.** Long-term mean surface temperatures for different eccentricities. Units are K. The color interval is 2 K. The black arrows show the migrations of the substellar points during an orbital cycle.

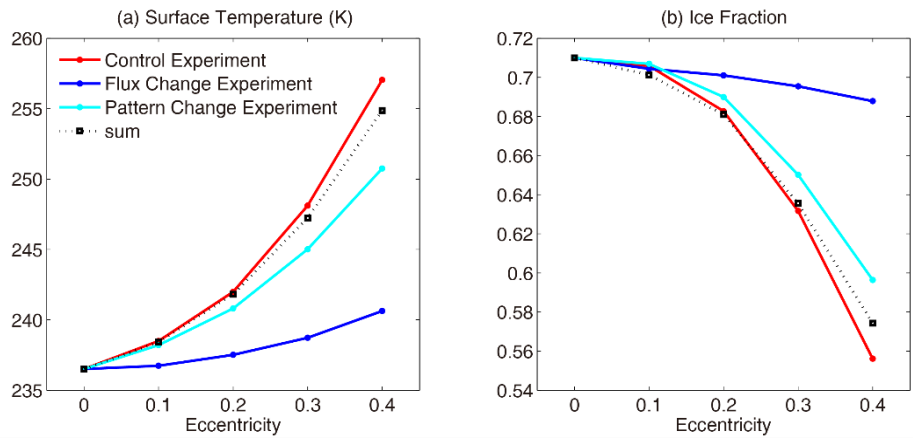

**Figure 5.** (a) Global mean surface temperatures for different eccentricities and different experiments. The red curve shows the total global mean surface temperature change caused by eccentricity. The deep blue curve shows the temperature change caused by annual mean insolation. The light blue curve shows the temperature change caused by insolation pattern. The black dotted curve shows the sum of the two blue curves. (b) The same with (a) but for ice fraction.

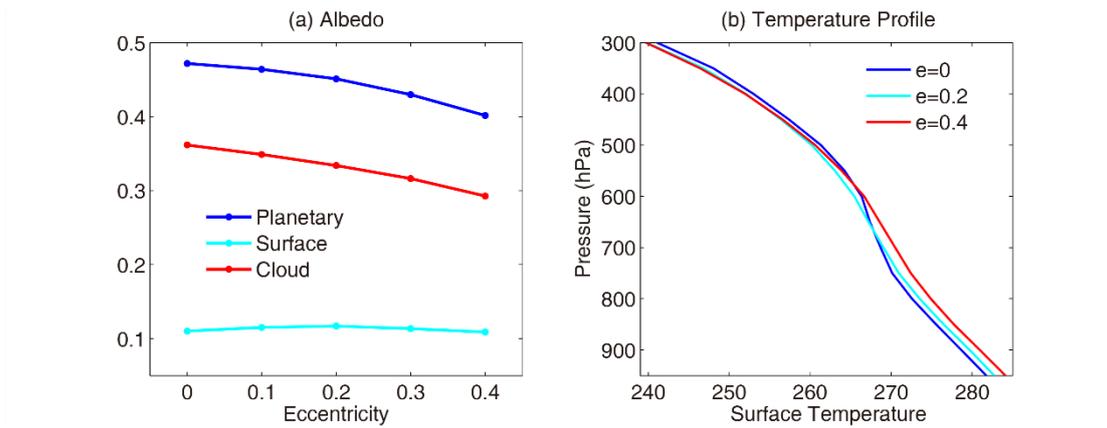

**Figure 6.** (a) Global mean planetary, surface and cloud albedos for different eccentricities in the "Pattern Change" experiments. (b) Temperature profiles averaged over permanent ocean regions for different eccentricities in the 'Control' experiments.

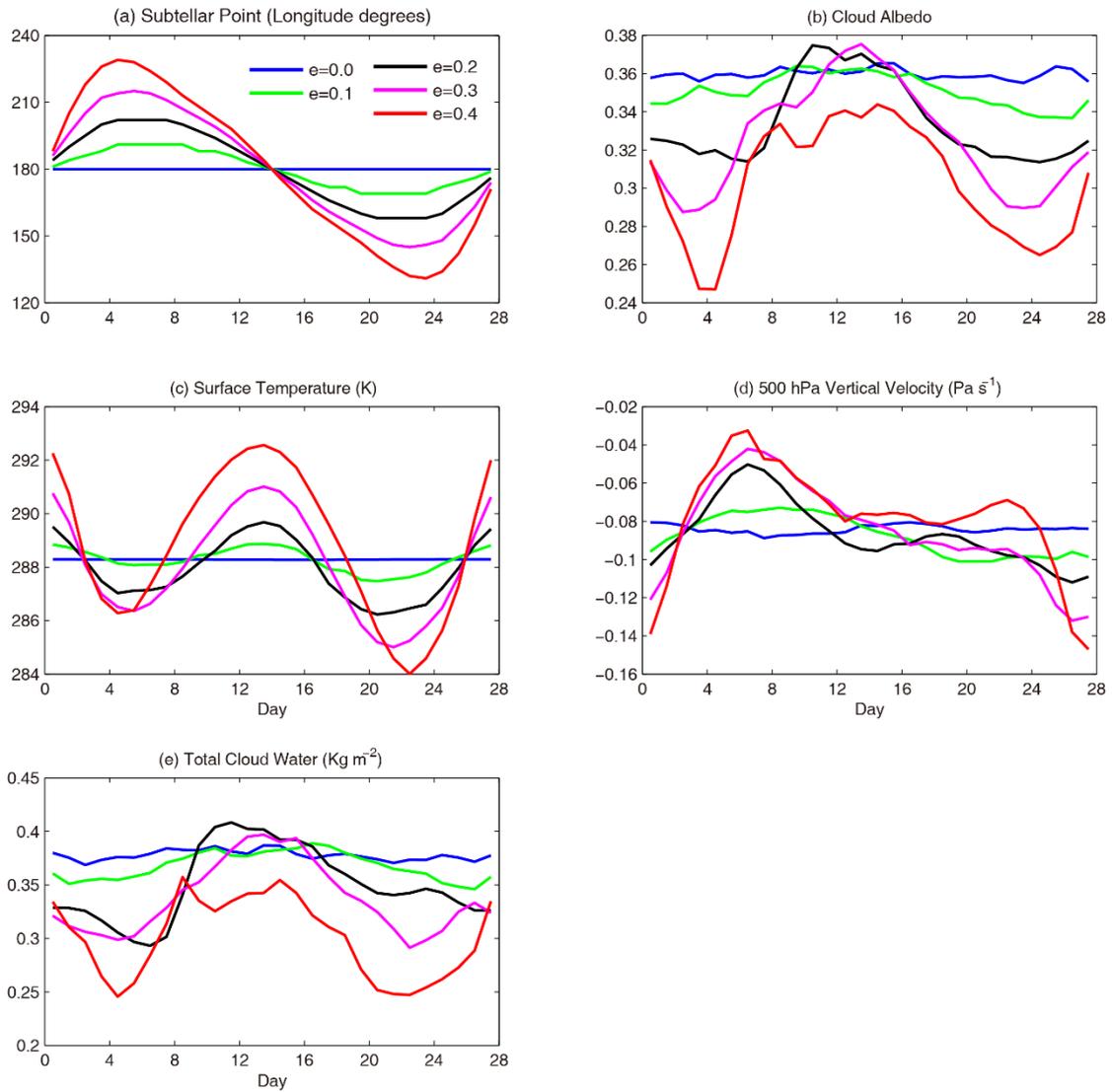

**Figure 7.** Variations of various quantities during one orbital cycle for different eccentricities in *p* =1.0 resonance state. The orbital period is 28 days. (a) Longitudinal position of the substellar point. (b) Cloud albedo. (c) Surface temperature, (d) vertical velocity at 500 hPa and (e) total cloud water. Quantities in (c) – (e) are averaged over a 60° radius circular disk around the substellar point.

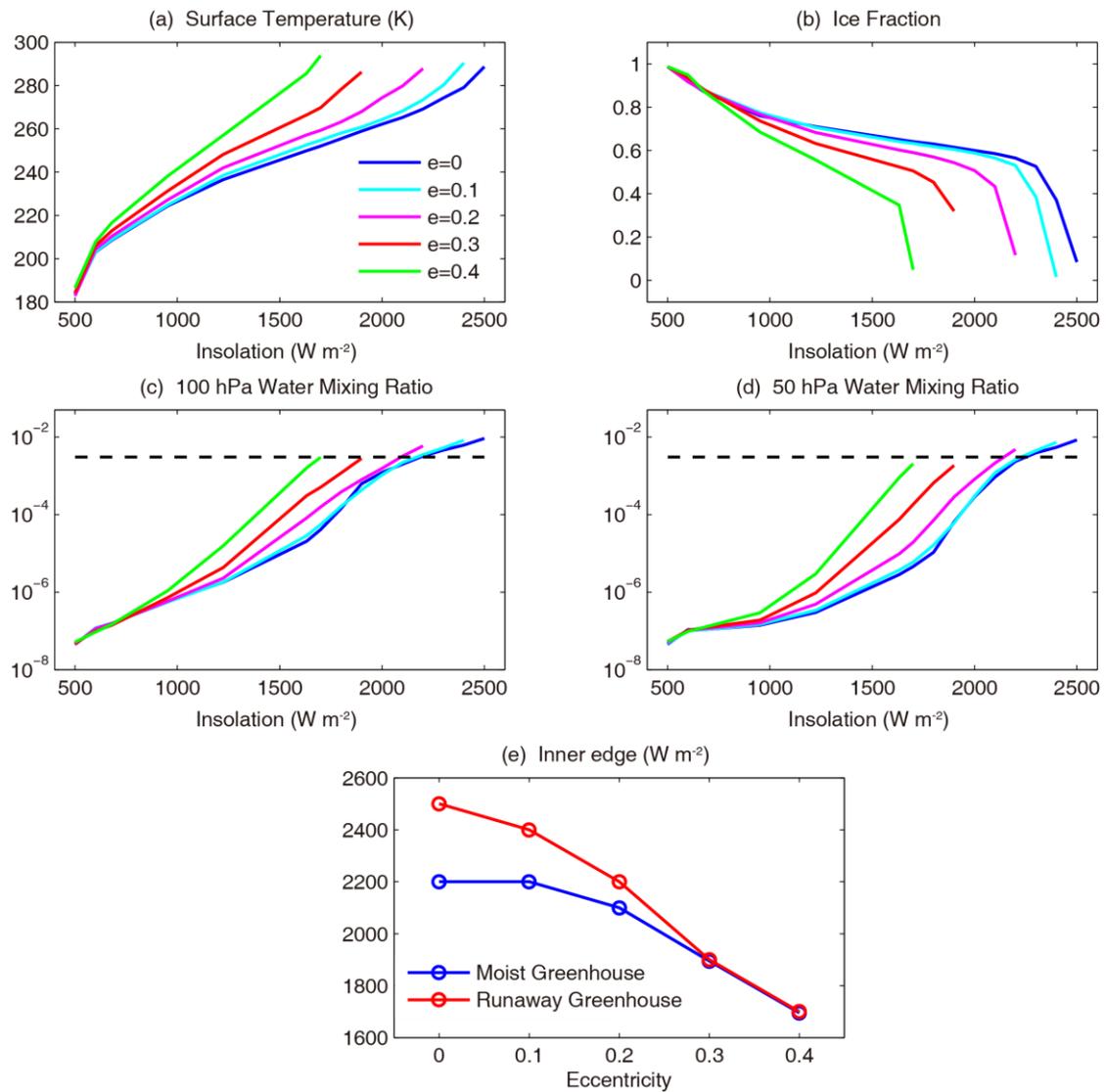

**Figure 8.** (a) Global mean surface temperature calculated for a series of insolations for each eccentricity ranging from 0 to 0.4 with a step of 0.1. (b) The same as (a) but for global mean ice fractions. (c) and (d) are the same as (a) but for water volume mixing ratio (including vapor, liquid and solid phases) at 100 hPa and 50 hPa. (e) Summary of the moist greenhouse and runaway greenhouse inner edges obtained from (a) and (c).

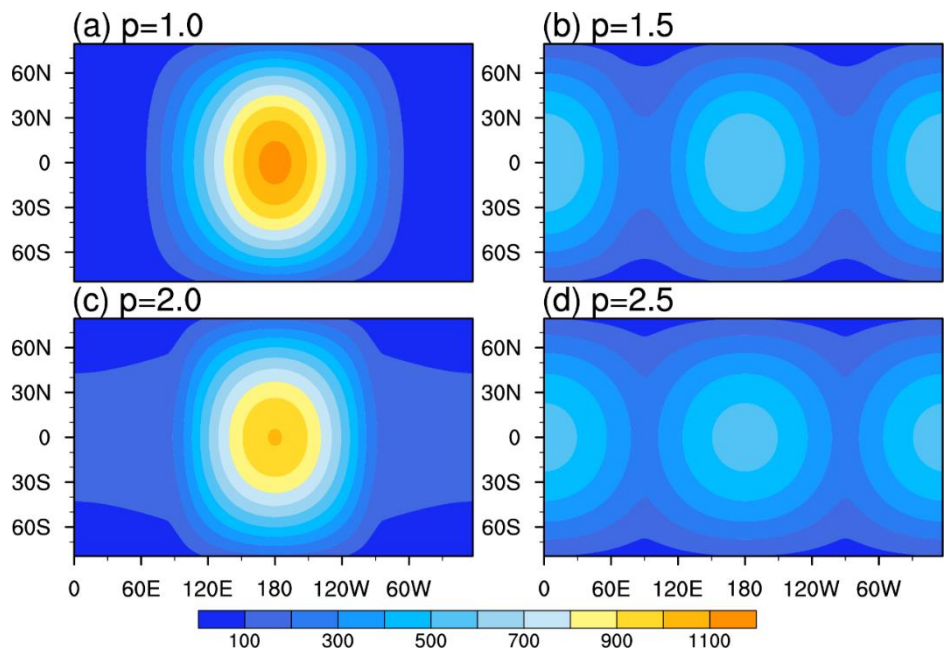

**Figure 9.** Incident stellar flux at the top of the atmosphere averaged over one orbital cycle for different spin-orbit resonance states. The eccentricity is 0.4. Contour interval is 100 W m$^{-2}$.

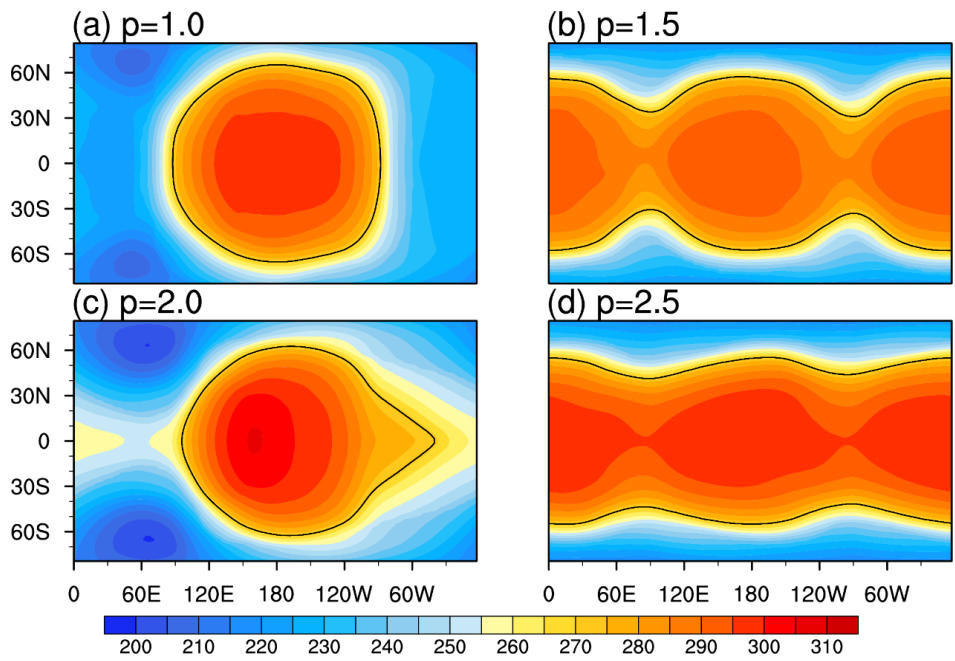

**Figure 10.** Annual mean surface temperature averaged over one orbital cycle for different spin-orbit resonance states. Contour interval is 5 K. The black curves show the isothermals of freezing point 271.35 K.

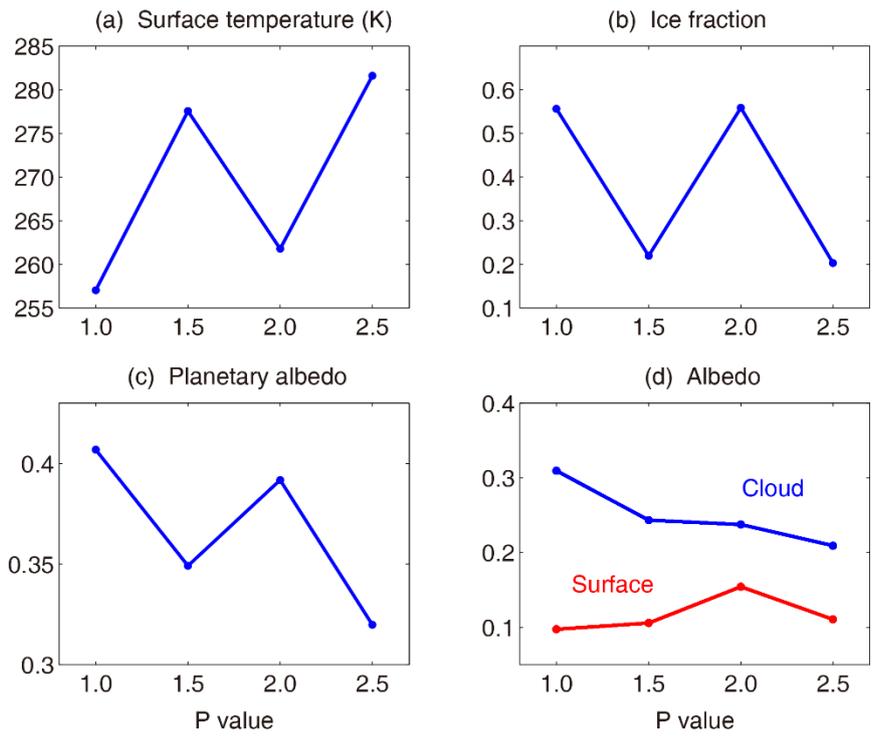

**Figure 11.** Climates of planets in different spin-orbit resonance states. (a) Global mean surface temperatures, (b) global mean ice fractions, (c) planetary albedos, and (d) cloud albedos and surface albedos.

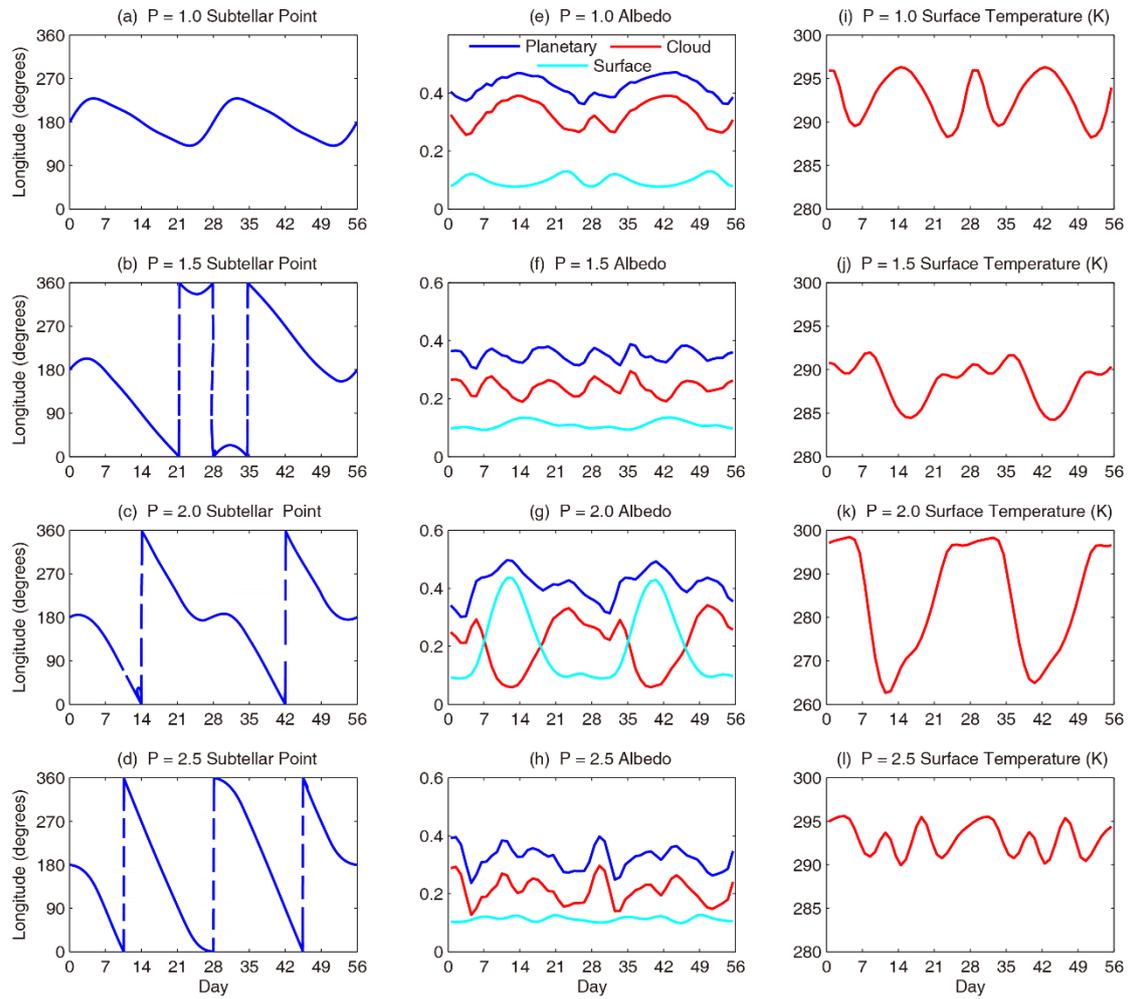

**Figure 12.** Variations of (left) locations of substellar point, (middle) global mean albedos and (right) surface temperature averaged over a 60° radius circular disk around the substellar point during two orbital cycles for different spin-orbit resonance states. The orbital period is 28 days. The eccentricity is 0.4.

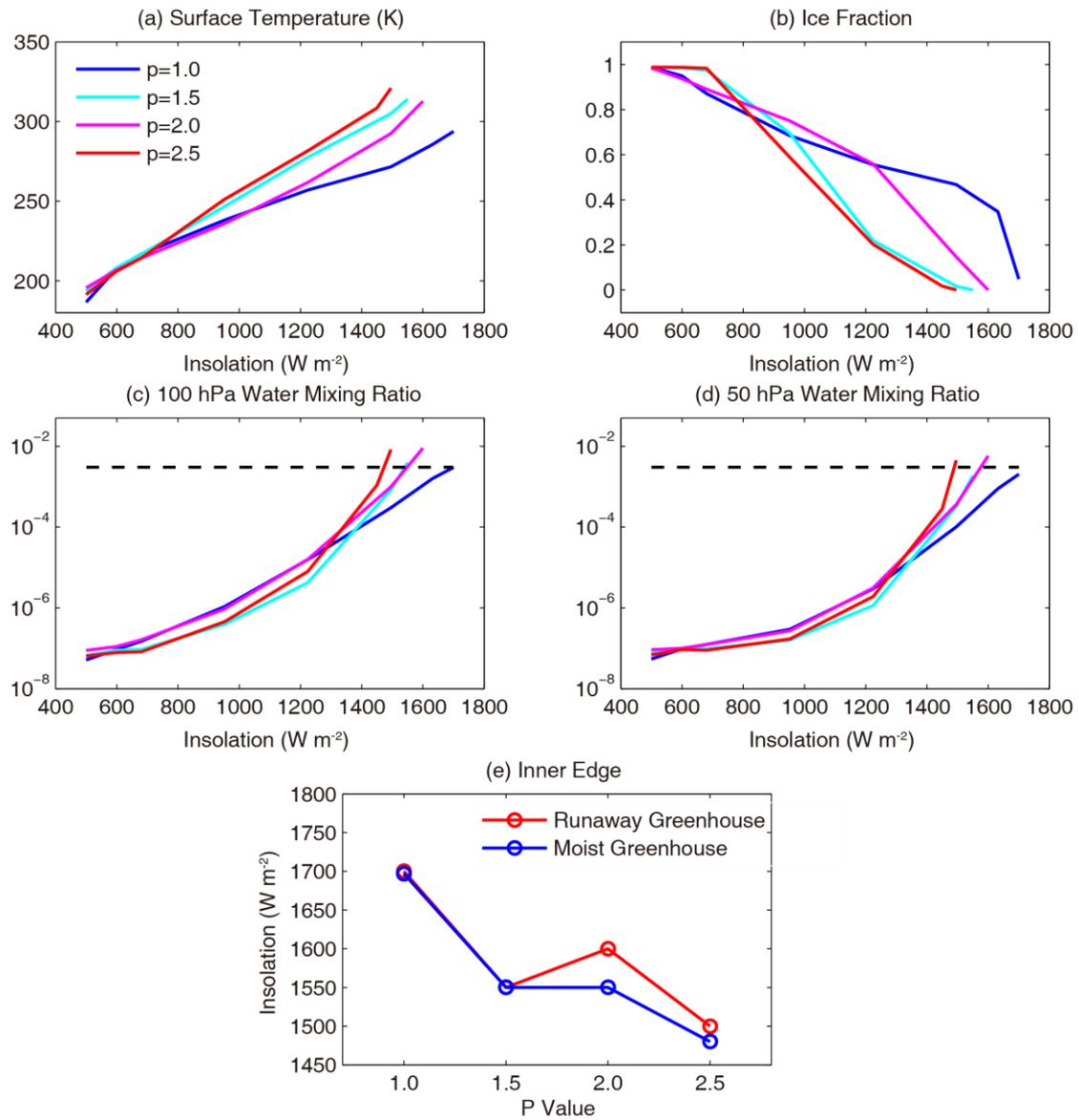

**Figure 13.** (a) Global mean surface temperature calculated for a series of insolations for spin-orbit resonances of 1.0, 1.5, 2.0 and 2.5. (b) The same as (a) but for global mean ice fractions. (c) and (d) are the same as (a) but for water volume mixing ratio at 100hPa and 50 hPa. (e) Summary of the moist greenhouse and runaway greenhouse inner edges obtained from (a) and (c).

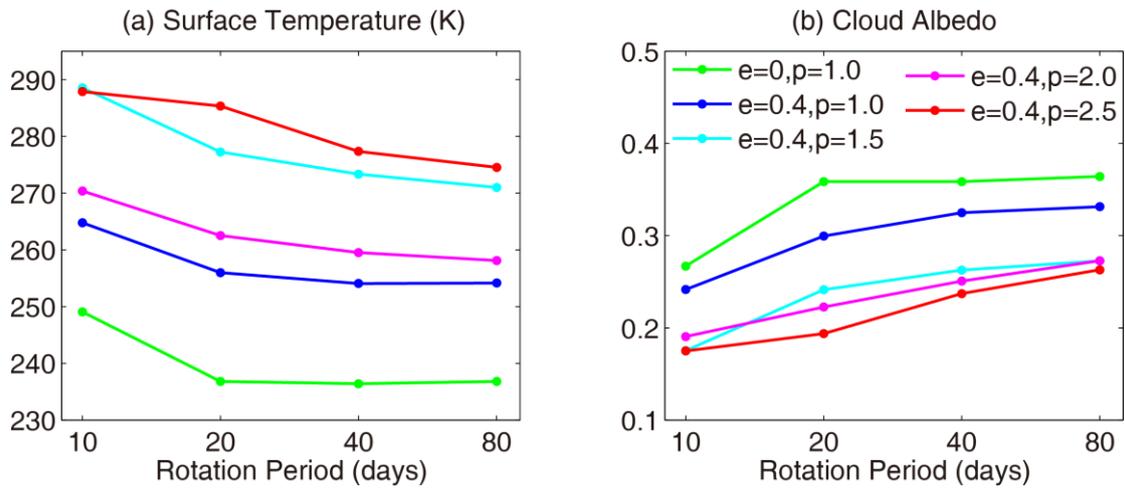

**Figure 14.** (a) Global mean surface temperatures for different eccentricities, spin-orbit resonance states, and rotational periods. (b) The same as (a) but for cloud albedos.